\documentclass[letterpaper,twocolumn,10pt]{article}
\usepackage{usenix}
\usepackage{subfigure}
\usepackage{fmtcount}
\usepackage{graphicx}
\usepackage{xspace}
\usepackage{url}
\usepackage{listings}
\usepackage{acronym}
\usepackage{algorithm}
\usepackage{latexsym}
\usepackage{fmtcount}
\usepackage{courier}
\usepackage{amsmath}

\DeclareMathOperator*{\argminA}{arg\,min}
\begin{document}
\title{\bf Explanation-Guided Diagnosis of Machine Learning Evasion Attacks}

\author{
{\rm Abderrahmen Amich}\\
University of Michigan, Dearborn\\
\texttt{aamich@umich.edu}
\and
{\rm Birhanu Eshete}\\
University of Michigan, Dearborn\\
\texttt{birhanu@umich.edu}
} 

\maketitle

\begin{abstract}
Machine Learning (ML) models are susceptible to evasion attacks. Evasion accuracy is typically assessed using aggregate evasion rate, and it is an open question whether aggregate evasion rate enables feature-level diagnosis on the effect of adversarial perturbations on evasive predictions.
In this paper, we introduce a novel framework that harnesses explainable ML methods to guide high-fidelity assessment of ML evasion attacks. Our framework enables explanation-guided correlation analysis between pre-evasion perturbations and post-evasion explanations. Towards systematic assessment of ML evasion attacks, we propose and evaluate a novel suite of model-agnostic metrics for sample-level and dataset-level correlation analysis. 
Using malware and image classifiers, we conduct comprehensive evaluations across diverse model architectures and complementary feature representations. Our explanation-guided correlation analysis reveals correlation gaps between adversarial samples and the corresponding perturbations performed on them. Using a case study on explanation-guided evasion, we show the broader usage of our methodology for assessing robustness of ML models.
\end{abstract}

\section{Introduction}\label{sec: intro}

The widespread usage of machine learning (ML) in a myriad of application domains has brought adversarial threats to ML models to the forefront of research towards dependable and secure ML systems. From image classification~\cite{ImageNet} to voice recognition~\cite{DL-Speech2012}, from precision medicine~\cite{DeepCC2019} to malware/intrusion detection~\cite{malconv18} and autonomous vehicles~\cite{DL-autnonmous17}, ML models have been shown to be vulnerable not only to training-time poisoning and evasion attacks, but also to model extraction and membership inference attacks~\cite{Wild-patterns18}. In typical evasion attacks, an adversary perturbs a legitimate input to craft an {\em adversarial sample} that tricks a victim model into making an incorrect prediction.

\textbf{Motivation:} Prior work has demonstrated adversarial sample-based evasion of ML models across diverse domains such as image classifiers~\cite{FGSM,BIM,CW,PGDM,Practical-black-box16}, malware classifiers~\cite{malconv18,MalGAN17,SLEIPINIR18,AdvApprox20,PDF-XuQE16,Android-DemontisMBMARCG19,GrossePMBM17,Apruzzese2020DeepRA}, and other domains such as speech and text processing~\cite{Carlini-list}. Evasion attacks have been explored across varying threat models (e.g., black-box ~\cite{MalConvEvade18,MalGAN17,Anderson-Reinf17,GrossePMBM17,APIBlackBoxEvade18,RNN-Black-Box18,ExploreAdvEx18,PDF-XuQE16}, white-box ~\cite{Android-DemontisMBMARCG19,ExploreAdvEx18,GANKnife18,SLEIPINIR18}). In the current state-of-the-art, the effectiveness of evasion is typically assessed through {\em aggregate evasion rate} by computing the percentage of crafted adversarial samples that lead a model to make evasive predictions.
For a ML model~$f$ that accepts a $d$-dimensional input~$x = [x_1,..., x_d]$ to predict~$f(x) = y_{true}$, the adversary perturbs~$x$ to obtain~$x' = [x_1+\delta_1,..., x_d+\delta_d]$, where~$\delta = [\delta_1,..., \delta_d]$ represents {\em pre-evasion perturbations} applied to each feature. When~$f$ is queried with~$x'$, it produces an evasive prediction~$f(x') = y' \ne y_{true}$. The natural question then is whether there exists {\em correlation between pre-evasion perturbations and the evasive prediction}. Unfortunately, aggregate evasion rate is inadequate to offer fine-grained insights to answer this question. In particular, it does not show how much the evasion strategy, through adversarial perturbations, influences individual samples to result in an evasive prediction. We consequently argue that unless one ``unpacks'' aggregate evasion rate at the resolution of an adversarial sample, it could give false sense of evasion success for it lacks the fidelity at the level of individual features. Such a coarse-grained nature of the aggregate evasion metric can potentially misguide the evaluation of model robustness in the face of adversarial manipulations.  

\textbf{Approach}: In this paper, we harness feature-based ML explanation methods and propose an explanation-guided correlation analysis framework for evasion attacks on ML models. Explainable ML techniques ~\cite{SHAP,LIME,LEMNA,DeepLIFT} interpret predictions returned by a ML model and attribute model's decision (e.g., predicted class label) to feature importance weights. In particular, for each evasive prediction $f(x') = y' \ne y_{true}$, explanation methods such as LIME~\cite{LIME} and SHAP~\cite{SHAP} produce {\em post-evasion explanations} of the form $[x_1: w_1, ..., x_d: w_d]$, where $w_i$ is the weight of contribution of feature $x_i$ to the evasive prediction $y'$. 
To address the lack of detailed insights from aggregate evasion rate, we leverage post-evasion explanations and empirically explore their feature-level correlations with pre-evasion perturbations performed by the adversary. Our key insight is that, since the perturbations are the only manipulations done on the feature-space of an input sample, when the model makes an evasive prediction on a perturbed variant of the input sample, there should exist some correlation between pre-evasion perturbations and post-evasion explanations. Towards systematic assessment of the link between adversarial perturbations and evasive predictions, we propose and evaluate a novel suite of metrics that allow (adversarial) sample-level and (evasion) dataset-level diagnosis of evasion attacks. Our suite of metrics are applicable to any ML model that predicts a class label given an input because, in the design of the metrics, we make no assumptions about the ML task and model architecture. The benefit of our fine-grained diagnosis for a defender whose aim is to evaluate ML model robustness to evasion attacks is twofold. First, it enables systematic measurement of the strength of correlation between an evasive prediction and feature-level perturbations across diverse classification tasks, model architectures, and feature representations. Second, it allows zooming-in on limitations of feature perturbation strategies to inform pre-deployment adversarial robustness evaluation of ML models.

 \textbf{Note on Scope}: We note that our approach is not yet another adversarial sample crafting approach for which problem-space to feature-space mapping is crucial to maintain functional integrity of adversarial samples (e.g., in adversarial malware samples). Our approach rather relies on feature-space perturbations performed to craft an adversarial sample and model output explanations of the same sample to perform correlation analysis and empirically examine the strength/weakness of the link between perturbations and explanations.

\textbf{Evaluation Highlights:} We evaluate our framework across different classification tasks (image, malware), diverse model architectures (e.g., neural networks, multiple tree-based classifiers, logistic regression),  and complementary  feature  representations (pixels for images, static and dynamic analysis-based features for malware). Our explanation-guided correlation analysis reveals an average of $45\%$ per-model adversarial samples that have low correlation links with perturbations performed on them --indicating the inadequacy of aggregate evasion rate, but the utility of fine-grained correlation analysis, for reliable diagnosis of evasion accuracy. Our results additionally suggest that, although a perturbation strategy evades a target model, at the granularity of each feature perturbation, it can lead to a per-model average of $36\%$ \textit{negative} feature perturbations (i.e., perturbations that contribute to maintain the original true prediction $f_b(x') = y_{true}$). We further evaluate the utility of our framework in a case study on explanation-guided adversarial sample crafting.

\noindent \textbf{Contributions:} In summary, this paper makes the following contributions:\\
$\bullet$ \textit{Explanation-guided diagnosis of evasion attacks.} To improve the sole reliance of evasion assessment on aggregate evasion rate, we propose an {\em explanation-guided correlation analysis} framework at the resolution of individual features. To that end, we introduce a novel {\em suite of correlation analysis metrics} and demonstrate their effectiveness at pinpointing adversarial examples that indeed evade a model, yet exhibit loose correlation with perturbations performed to craft them.\\
$\bullet$ \textit{Comprehensive evaluations.} In malware classification and image classification, we conduct extensive evaluations across diverse models architectures and feature representations, and synthesize interesting experimental insights that demonstrate the utility of explanation-guided correlation analysis.\\
$\bullet$ \textit{Further case study.} We conduct a case study that demonstrates a practical use-case of our framework via \textit{pre-perturbation feature direction analysis} to guide evasion strategies towards crafting more accurate adversarial samples \textit{correlated} with their evasive predictions.

\section{Background: ML Evasion and Explanation Methods}\label{sec: bground}
In this section, we succinctly introduce ML evasion attacks and ML explanation methods.



\subsection{ML Evasion Attacks}\label{subsec:test-time}

\textbf{Adversarial Sample Crafting.} Given a deployed ML model (e.g., malware classifier, image classifier) with  a decision function $f:X \rightarrow Y$ that maps an input sample $x \in X$ to a true class label $y_{true} \in Y$, then $x'$ = $x  + \delta$ is called an {\em adversarial sample} with an {\em adversarial perturbation} $\delta$ if: 
    $f(x') = y' \ne y_{true}, ||\delta|| < \epsilon$,
where $||.||$ is a distance metric (e.g., one of the $L_{p}$ norms)  and $\epsilon$ is the maximum allowable perturbation that results in misclassification while preserving semantic integrity of $x$. Semantic integrity is domain and/or task specific. For instance, in image classification, visual imperceptibility of $x'$ from $x$ is desired while in malware detection $x$ and $x'$ need to satisfy certain functional equivalence (e.g., if $x$ was a malware pre-perturbation, $x'$ is expected to exhibit maliciousness post-perturbation as well). In {
\em untargeted} evasion, the goal is to make the model misclassify a sample to any different class (e.g., for a roadside sign detection model: misclassify red light as any other sign). When the evasion is {\em targeted}, the goal is to make the model to misclassify a sample to a specific target class (e.g., in malware detection: misclassify malware as benign).

Evasion attacks can be done in {\em white-box} or {\em black-box} setting. Most gradient-based evasion techniques ~\cite{FGSM,BIM,PGDM,CW} are white-box because the adversary typically has access to model architecture and parameters/weights, which allows to query the model directly to decide how to increase the model’s loss function. In recent years, several white-box adversarial sample crafting methods have been proposed, specially for image classification tasks. Some of the most notable ones are: Fast Gradient Sign Method (FGSM)~\cite{FGSM}, Basic Iterative Method (BIM)~\cite{BIM}, Projected Gradient Descent (PGD) method~\cite{PGDM}, and Carlini \& Wagner (CW) method~\cite{CW}. Black-box evasion techniques usually start from some initial perturbation $\delta_{0}$, and subsequently probe $f$ on a series of perturbations $f (x + \delta_{i})$, to craft a variation of $x$ that evades $f$ (i.e., misclassified to a label different from its original). 
In malware classifiers, while {\em gradient-based} methods have been widely adopted both in white-box and black-box settings, two other strategies also standout for evasion in a black-box setting. The first one is called {\em additive} because it appends adversarial noise (e.g., no-op bytes) to the end of a sample (e.g., Windows PE) in order to preserve original behavior~\cite{EndtoEnd18,GrossePMBM17}. The second class of methods uses {\em targeted and constrained manipulations} after identifying regions in the PE that are unlikely to be mapped to memory~\cite{ExploreAdvEx18}. 

 One of the challenges for state-of-the-art adversarial sample crafting methods is the lack/impossibility of mapping of feature-space perturbations to the problem space. Such a mapping and reversibility between the two spaces is essential in domains where the functionality of an adversarial sample needs to be preserved post-perturbation~\cite{Problem-SpaceNIDS-20}. A recent work by Pierazzi et al.~\cite{ProblemSpace-20} proposes formulations and shows promising experimental results towards the feasibility of crafting evasive malware samples with real-world consequences. As explained in section \ref{sec: intro}, we recall that problem space perturbations are out of this paper's scope.

\subsection{ML Explanation Methods}\label{subsec:attribution} 
Humans typically justify their decision by explaining underlying causes used to reach a decision. For instance, in an image classification task (e.g., cats vs. dogs), humans attribute their classification decision (e.g., cat) to certain parts/features (e.g., pointy ears, longer tails) of the image they see, and not all features have the same importance/weight in the decision process. ML models have long been perceived as black-box in their predictions until the advent of explainable ML~\cite{LIME,DeepLIFT,SHAP}, which attribute a decision of a model to features that contributed to the decision. This notion of attribution is based on quantifiable contribution of each feature to a model's decision.

ML explanation is usually accomplished by training a substitute model based on the input feature vectors and output predictions of the model, and then use the coefficients of that model to approximate the importance and {\em direction} (class label it leans to) of the feature.  A typical substitute model for explanation is of the form:
    $s(x) = w_{0} +\sum_{i=1}^{d} w_{i}x_{i}$,
where $d$ is the number of features, $x$ is the sample, $x_{i}$ is the $i^{th}$ feature for sample $x$, and $w_{i}$ is the contribution/weight of feature $x_{i}$ to the model’s decision. While ML explanation methods exist for white-box \cite{Whitebox-exp13,Whitebox-exp14} or black-box~\cite{LIME,LEMNA,SHAP} access to the model, in this work we consider ML explanation methods that have black-box access to the ML model, among which the notable ones are LIME \cite{LIME}, SHAP \cite{SHAP} and LEMNA \cite{LEMNA}. Next, we briefly introduce these explanation methods.

\textbf{LIME and SHAP.}\label{LIME_SHAP}
Ribeiro et al. \cite{LIME} introduce LIME as one of
the first model-agnostic black-box methods for locally explaining model output. Lundberg and Lee further extended LIME by proposing SHAP \cite{SHAP}. Both methods approximate the decision function $f_b$ by creating a series of $l$ perturbations of a sample $x$, denoted as $x'_1, . . . , x'_l$ by randomly setting feature values in the vector $x$ to $0$. The methods then proceed by predicting a label $f_b(x'_i)=y_i$ for each $x'_i$ of the $l$ perturbations. This sampling strategy
enables the methods to approximate the local neighborhood of $f_b$ at the point $f_b (x)$. LIME approximates the decision boundary by a weighted linear regression model using Equation \ref{eq:LIME}.
\begin{equation}\label{eq:LIME}
   \argminA_{g \in G} \sum_{i=1}^{l} \pi_x(x'_i)(f_b(x'_i)-g(x'_i))^2
\end{equation}
In Equation \ref{eq:LIME}, $G$ is the set of all linear functions and $\pi_x$ is a function indicating the difference between the input $x$ and a perturbation $x'$. SHAP follows a similar approach but employs the SHAP kernel as weighting function $\pi_x$, which is computed using the \textit{Shapley Values} \cite{shapley} when solving the regression. Shapley Values are a concept from game theory where the features act as players under the objective of finding a fair contribution of the features to the payout --in this case the prediction of the model.

\textbf{LEMNA.}\label{LEMNA}
Another black-box explanation method specifically designed to be a better fit for non-linear models is LEMNA \cite{LEMNA}. As shown in Equation \ref{eq:LEMNA}, it uses a mixture regression model for approximation, that is, a weighted sum of $K$ linear models.
\begin{equation}\label{eq:LEMNA}
    f(x)= \sum_{j=1}^{K} \pi_j(\beta_j . x +\epsilon_j)
\end{equation}
In Equation \ref{eq:LEMNA}, the parameter $K$ specifies the number of models, the random variables $\epsilon = (\epsilon_1, ... , \epsilon_K)$ originate from a normal distribution $\epsilon_i \sim N(0, \sigma)$ and $\pi = (\pi_1, ... , \pi_K)$ holds the weights for each model. The variables $\beta_1, ... , \beta_K$ are the regression coefficients and can be interpreted as $K$ linear approximations of the decision boundary near $f_b(x)$.
\section{Explanation-Guided Evasion Diagnosis Framework}\label{sec: approach}
In this section, we present our explanation-guided correlation analysis methodology. Table \ref{tab:notations} describes notations used here and in the rest of the paper.

\begin{table*}[t!]
\caption{Notations.}\label{tab:notations}
 \begin{center}
   \begin{tabular}{|r|l|} 
      \hline
      \textbf{Notation} & \textbf{Brief Description} \\ \hline
       $X_b$ & training set of black-box model $f_{b}$. \\
       $X_e$ & evasion set disjoint with $X_b$. \\
      $X'_e$ & adversarial counterpart of $X_e$. \\
      $X_{s}$ & training set of explanation model $f_{s}$. \\
      $x \in X_e$ & sample in evasion set. \\
       $x' \in X'_e$ & adversarial variant of $x$. \\
       $ Y = \{y_1,...,y_k\}$& set of classes (labels)\\
    
        $x= [x_1,...,x_d]$& $d$-dimensional feature vector of sample $x$. \\
       $W_{x,y_i}= \{w_1,...,w_d\}$& feature weights (explanations) of a sample $x$ toward the class $y_i$. \\
       $pos(x,y_i)$ & number of features in $x$ {\em positive} to the prediction $f_b(x)=y_i$\\
       $neg(x,y_i)$ & number of features in $x$ {\em negative} to the prediction $f_b(x)=y_i$\\
       $neut(x,y_i)$ & number of features in $x$ {\em neutral} to the prediction $f_b(x)=y_i$\\
       $P(x')$ & number of perturbed features in $x'$\\
       $\tau$ & threshold to decide highly-correlated adversarial samples. \\ \hline
       
   \end{tabular}
 \end{center}
\end{table*} 

\subsection{Overview}

As described in Section \ref{sec: intro}, the effectiveness of an evasion method is typically assessed using aggregate evasion accuracy. While aggregate evasion quantifies the overall success of an evasion strategy, it fails to offer sufficient insights. In particular, it does not show how the evasion mechanism influences individual samples to result in evasive predictions. We argue that, unless one examines evasion success at the resolution of each adversarial sample, aggregate evasion rate could give false sense of adversarial success for it lacks feature-level fidelity of perturbations that result in an adversarial sample. To address the stated lack of fidelity in aggregate evasion accuracy, we systematically explore how ML explanation methods are harnessed to assess feature-level correlations between pre-evasion adversarial perturbations and post-evasion explanations. 

\begin{figure*}[t!]
    \centering
    \includegraphics[width=\textwidth]{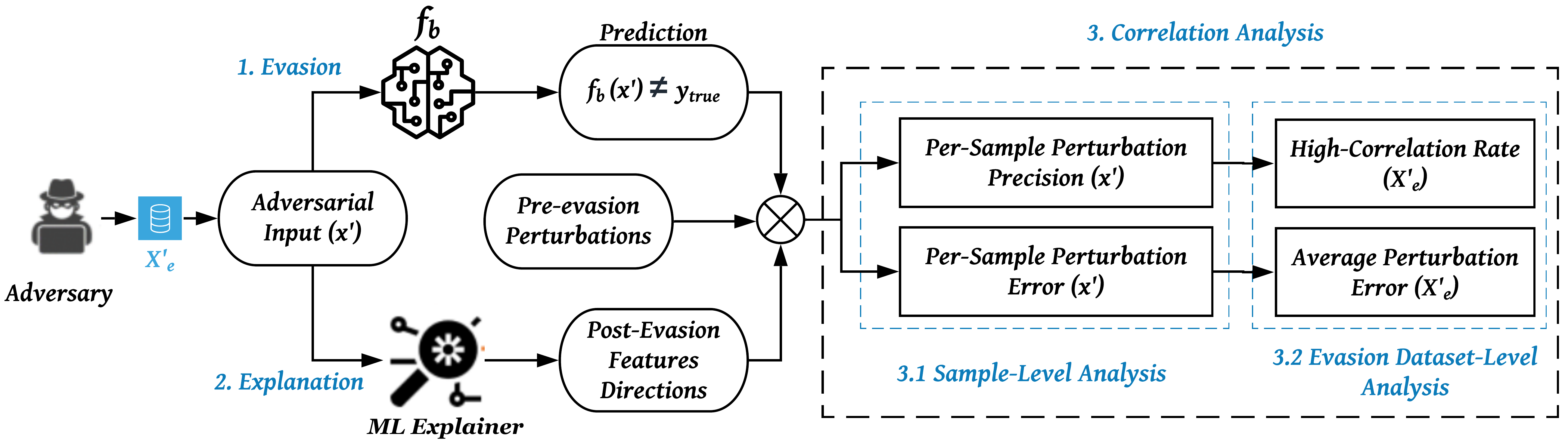}
    \caption{Explanation-guided correlation analysis framework.}
    \label{fig:framework}

\end{figure*}

Figure \ref{fig:framework} shows an overview of our explanation-guided correlation analysis framework. Given an evasion set $X'_e$ of adversarial samples, our framework enables correlation analysis both at the sample-level (for each $x' \in X'_e$ at the granularity of each perturbed feature) and at the evasion dataset-level ($ \forall x' \in X'_e$). Intuitively, given a decisive feature (obtained via ML explanations) of an evasive sample ($f_b(x') \neq y_{true}$), for such a feature to be considered the cause of (correlated to) the evasion, there needs to be a corresponding feature that was perturbed in the original sample $x$. By repeating the correlation of each decisive feature with its perturbed counterpart, our sample-level correlation analysis establishes empirical evidence that links an evasive prediction with its cause.  

More precisely, our correlation analysis is performed by harnessing the \textit{post-evasion features directions} (``2. Explanation'' in Figure \ref{fig:framework}) of adversarial samples (``1. Evasion'' in Figure \ref{fig:framework}). First, we explore the feature directions of the {\em pre-evasion perturbations} in order to obtain an assessment of the contribution of each feature perturbation to the attack (i.e., \textit{feature-level assessment}). Second, we use those results in order to zoom-out to a \textit{sample-level assessment} (\ref{subsec:samp_ana}). Finally, we move to the higher level of the whole evasion dataset in order to obtain an overall assessment of the evasion attack (\ref{subsec:data_ana}).

Conducting such fine-grained correlation analysis has two key benefits. Firstly, it verifies whether evasion can be attributed to the adversarial perturbations employed on the sample, and, in effect, performs diagnosis on aggregate evasion accuracy. Secondly, it provides visibility into how sensitive certain samples and/or features are to adversarial perturbations, which could inform robustness assessment of ML models in the face of evasion attacks.

\subsection{Post-Evasion Feature Direction}\label{subsec:feat_dir}

In a typical classification task, for an input sample $x$,  $f_b (x) = y_i \in Y=\{y_1,...,y_k\}$, where $Y$ is the set of $k$ possible labels. For example, in the multi-class handwritten digit recognition model of the MNIST \cite{MNIST} dataset, the input is an image of a handwritten digit and the label is one of the 10 digits (i.e., $Y = \{0,..,9\}$ where $k=10$). In the malware detection domain, the typical model is a binary classifier (i.e., $Y = \{{\tt Benign}, {\tt Malware}\}$ where $k=2$). Next, we use MNIST as an illustrative example to describe post-evasion feature direction.

Explanations returned from ML explanation methods reveal the {\em direction} of each feature. For each class $y_i \in Y$ and an adversarial sample $x'$, a ML explanation method returns a set of {\em feature weights} $W_{x',y_i} = \{w_1,..,w_d\}$ where $w_j$ reflects the importance (as the magnitude of $w_j$) and the {\em direction} (as the sign of $w_j$) of the feature $x'_j$ towards the prediction $f_b(x')=y_i$. Depending on the sign of $w_j$, feature $x'_j$ can be {\em positive}, {\em negative}, or {\em neutral} with respect to the prediction $f_b(x')=y_i$. When $w_j>0$, we say $x'_j$ is positive to (directed towards) $y_i$. Conversely, when $w_j<0$, $x'_j$ is negative to (directed away from) $y'_i$. When $w_j=0$, we say $x'_j$ is neutral to $y_i$ (does not have any impact on the prediction decision). In case of binary classification ($k=2$), if $x'_j$ is not directed to the label $y_1$ ({\tt Benign} for malware detection) and is not neutral, then $x'_j$ can only be directed to the other label $y_2$ ({\tt Malware}) and vice versa.  To illustrate how we leverage feature direction in our analysis, next we describe a concrete example from the MNIST \cite{MNIST} handwritten digit recognition model.

\begin{figure*}[t!]

    \centering
    \includegraphics[width=\textwidth]{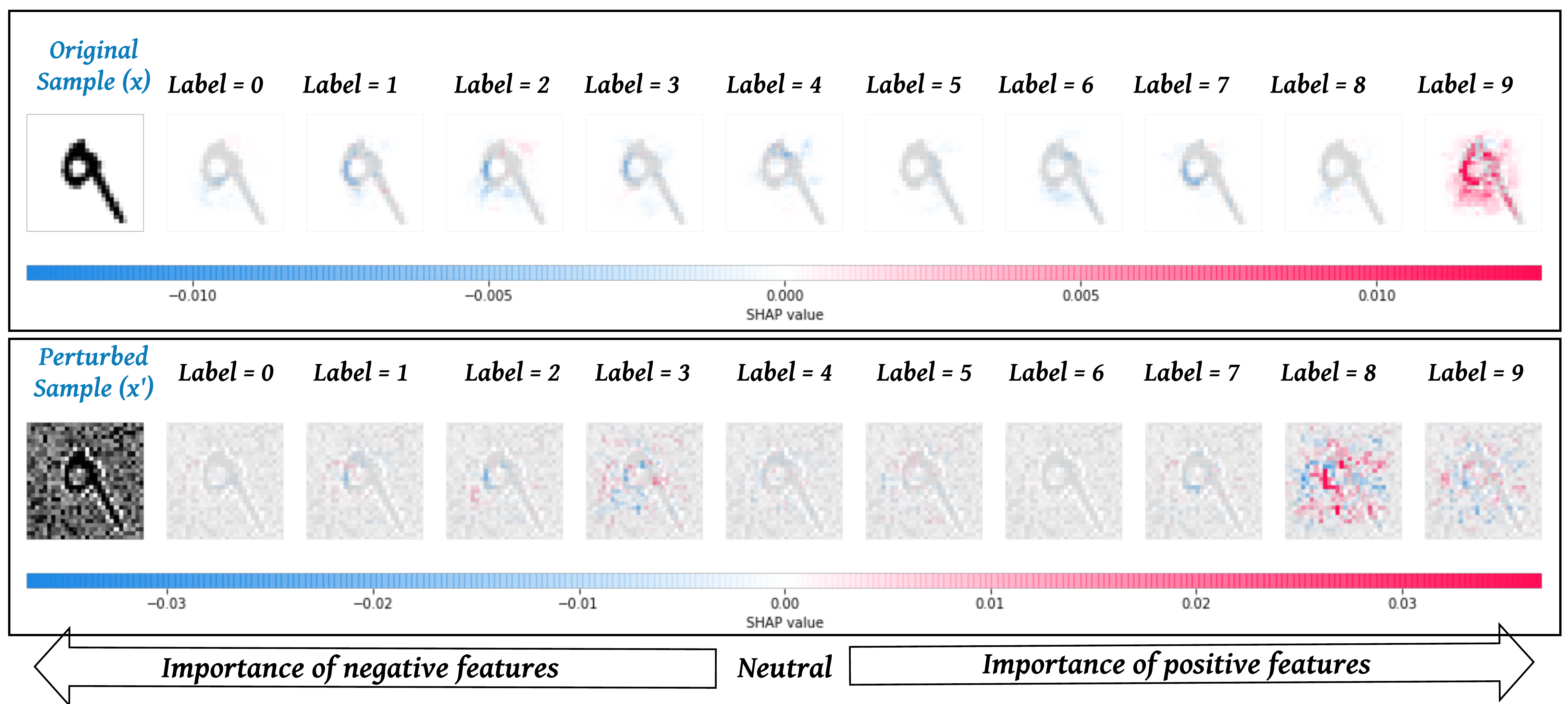}
    \caption{A comparative illustration of pre-perturbation and post-perturbation explanations using the SHAP \cite{SHAP} ML explanation framework on a test sample from the MNIST \cite{MNIST} dataset.}
    \label{fig:org_exp}

\end{figure*}

In Figure \ref{fig:org_exp}, the upper box shows SHAP \cite{SHAP} pre-evasion feature explanations of a correct prediction on an image of ``9`` (i.e., $f_b(x)=9$). The lower box shows post-evasion feature explanations of the misclassification $f_b(x')\neq9$ using an adversarial variant $x' \gets x+\delta$. Each column (i.e., ``$Label = y_i$``; $y_i \in \{0,...,9\}$) represents the feature directions for the possibility of a prediction $f_b(x) = y_i$ (upper box) and $f_b(x')=y_i$ (lower box). The color codes are interpreted as follows: given an explanation, pink corresponds to positive features while blue corresponds to negative features. The intensity of either color (pink or blue) is directly proportional to the feature weight towards the prediction. Neutral features are represented with white.  For instance, focusing on the correct prediction label $y_{true} = 9$ in the upper-box, we notice a large concentration of pink features which positively contribute to the predicted label ($9$).  

Our approach primarily relies on post-evasion explanations (lower box in Figure \ref{fig:org_exp}) and we observe that feature importance weights vary for each studied label as a potential prediction $f_b(x') = y_i \in \{0,...,9\}$. When the prediction $f_b(x') = 8$ (image below 'Label = 8' in lower box), the explanations show that most features are \textit{positive} (directed to label $8$), which explains the change of the prediction label from $9$ to $8$. Examining the colors, we realize that most features that were directed to label $9$ in the pre-perturbation explanations have become either neutral to the prediction $f_b(x') = 9$ or are positive towards $f_b(x') = 8$. It is noteworthy that some perturbed features are oriented to the original label $9$ (notice pink pixels in the image below 'Label = 9' in lower box). Such observations suggest that even though the attack is successful (i.e., $f_b(x') = 8 \neq 9$), the effectiveness of each single feature perturbation is not guaranteed to result in an evasive prediction. Thus, the evasion success may not always be correlated with each feature perturbation the adversary performs on the original sample. We, therefore, argue that a perturbation strategy that produces many features that are uncorrelated with the misclassification might perform poorly on other feature representations (e.g., colored or not centered images in image classification) or other feature types (e.g., static vs. dynamic features in malware detection) which reflects a potential limitation of the stability of a perturbation method.
Next, we introduce novel sample-level metrics that capture the fine-grained assessment that leverages post-evasion explanations. We refer to Table \ref{tab:notations} for the feature direction-related notations. Our focus will be on the \textit{post-perturbation feature directions} of an evasive sample $x'$ and we suppose that its original prediction (pre-perturbation) is $f_b(x) = y_{true}$.

\subsection{Sample-Level Analysis}\label{subsec:samp_ana}

Post-evasion explanations reveal the direction (\textit{positive}, \textit{negative}, or \textit{neutral}) of each perturbed feature in an adversarial sample $x'$. Sample-level analysis is performed in order to empirically assess feature perturbations that positively contribute towards misclassification (\textit{positive perturbations}) against the ones that contribute to maintain the true label as a prediction (\textit{negative perturbations}). Next, we introduce two sample-level metrics which will later serve as foundations to conduct overall correlation analysis over the evasion dataset. 

\textbf{Definition 1: Per-Sample Perturbation Precision (PSPP).}
Out of all performed feature perturbations ($P(x')$) to produce an adversarial sample $x'$, \textit{PSPP} enables us compute the rate of perturbations that contribute to change the original prediction $y_{true}$ to another label $y_i\in Y-\{y_{true}\}$. In other words, it measures the rate of perturbed features that are \textit{``negative"} to the original prediction ($y_{true}$) and \textit{``positive"} to other predictions $y_i \neq y_{true}$. We call such perturbations \textit{positive perturbations} because they positively advance the evasion goal. More formally, the Per-Sample Perturbation Precision for an adversarial sample $x'$ is computed as follows:
\begin{equation}\label{eq:PP}
    PSPP(x') =\frac{1}{2}(\frac{1}{k-1}(\sum_{\substack{y_i \in Y \\ y_i \neq y_{true}}} \frac{pos(x',y_i)}{P(x')} )+ \frac{neg(x',y_{true})}{P(x')})
\end{equation} 

Equation \ref{eq:PP} is the average of two ratios:\\
  $\bullet$ ($\frac{1}{k-1}(\sum_{\substack{y_i \in Y \\ y_i \neq y_{true}}}^{} \frac{pos(x',y_i)}{P(x')}$): The average rate of perturbed features that are directed to a class $y_i \neq y_{true}$, over all $k-1$ possible false classes $y_i\in Y-\{y_{true}\}$.\\
  $\bullet$ ($\frac{neg(x',y_{true})}{P(x')}$): The rate of perturbed features that are not directed to the original label $y_{true}$ and not neutral.

Both ratios that are considered in Equation \ref{eq:PP} measure \textit{Positive Perturbations} that contribute to a misclasssification. We note that $PSPP(x')$ falls in the range $[0,1]$. The closer $PSPP(x')$ is to $1$, the more the overall perturbations performed on the features of $x'$ are precise (effective at feature level). More importantly, when $x'$ evades the model, i.e., $f_b(x') \neq f_b(x)$, then the closer $PSPP(x')$ is to 1 the stronger the correlation between the evasion success and each performed feature perturbation that produced adversarial sample $x'$.

\textbf{Definition 2: Per-Sample Perturbation Error (PSPE).}
Another per-sample measurement for our correlation analysis is the Per-Sample Perturbation Error, $PSPE(x')$, that computes the rate of perturbed features that are directed to the original class $y_{true}$ (\textit{positive} to the original prediction $y_{true}$). These features stand against the adversary's goal of misclassifying $x'$. Such features are considered \textit{negative perturbations} with respect to the original class. More formally, $PSPE(x')$ is defined as follows:
\begin{equation}\label{eq:PSPE}
    PSPE(x') =\frac{pos(x',y_{true})}{P(x')}
\end{equation} 
Given an adversarial sample $x'$, $PSPE(x')$ returns the rate of perturbation errors over all perturbed features. We note that a perturbed feature that is \textit{neutral} ($w_j = 0$) to the original prediction ($f_b(x') = y_{true}$) is considered neither as perturbation error nor an effective manipulation to advance the evasion goal. Thus, $PSPE(x')$ may not be directly computed from $PSPP(x')$ and vice versa. Moreover, in the case of a slightly different threat model in-which the evasion is \textit{targeted} to change the original prediction $y_{true}$ to a new target label $y_{target} \in Y-\{y_{true}\}$, then only the term $\frac{pos(x',y_{target})}{P(x')}$ would be considered to compute the perturbation precision $PSPP(x')$, and only the term $\frac{neg(x',y_{target})}{P(x')}$ suffices to compute the rate of committed perturbation errors, $PSPE(x')$.


\subsection{Evasion Dataset-Level Analysis}\label{subsec:data_ana}
Using $PSPP (x')$ and $PSPE(x')$ defined in Equations \ref{eq:PP} and \ref{eq:PSPE} as foundations, we now introduce novel correlation analysis metrics that operate at the level of the evasion dataset $X'_e$ to empirically analyze correlation between perturbations and post-evasion explanations.

\textbf{Definition 3: High-Correlation Rate (HCR).}\label{sec:HCR}
As explained in Section \ref{subsec:samp_ana}, $PSPP(x')$ quantifies the correlation of each single feature perturbation with the evasion $f_b(x') \neq y_{true}$. The closer $PSPP(x')$ is to $1$, the higher is the correlation and vice-versa. We consider a threshold $\tau$ that indicates the ``strength'' of the correlation between positive perturbations on $x$ that resulted in $x'$ and the important features that ``explain'' $f_b(x') \neq y_{true}$. Based on an empirically estimated $\tau$, we call an adversarial sample $x'$ a \textit{High-Correlated Sample} if $PSPP(x')$ falls in $[\tau,1]$. In our evaluation, based on empirical observations, we use $\tau=0.5$.

Based on the above definition, we compute \textit{High-Correlation Rate (HCR)} as the percentage of \textit{High-Correlated Samples} in the evasion set $X'_e$ as follows:
    
\begin{equation}\label{eq:TER}
    HCR=\frac{|X_e'(PSPP>\tau)|}{|X_e'|} 
\end{equation} 
where \resizebox{0.9\hsize}{!}{$X_e'(PSPP>\tau)=\{x' \in X'_e:\,PSPP(x')>\tau\} \cap \{f_b(x') \neq y_{true}\}$}.
We note that {\em HCR} quantifies the degree to which adversarial samples are both \textit{evasive} and \textit{correlated} to most feature perturbations performed on original samples.

\textbf{Definition 4: Average Perturbation Error (APE).}\label{sec:APE}
As shown in Equation \ref{eq:PSPE}, $PSPE(x')$ computes the number of errors committed during the perturbation of each feature in $x$ to produce the manipulated sample $x'$ (which is the same as computing the number of \textit{negative perturbations}). We leverage $PSPE(x')$ to compute the average of negative perturbations ($APE$) over all samples in $X'_e$. Formally, $APE$ is given as follows:
\begin{equation}\label{eq:APE}
    APE= \sum_{x' \in X_e'}^{} \frac{PSPE(x')}{|X_e'|}
\end{equation} 

As opposed to aggregate evasion rate that computes the percentage of \textit{evasive samples} versus \textit{non-evasive samples} without deeper insights about the effectiveness of each single feature perturbation, $APE$ computes the rate of \textit{``evasive features"} versus \textit{``non-evasive features"} of each evasive sample, over all perturbed samples. Such in-depth investigations into evasion attacks provide a fine-grained assessment of any evasion strategy on ML models.
\section{Evaluation}\label{sec: eval}
We now evaluate the utility of our suite of metrics for high-fidelity correlation analysis of ML evasion attacks. We first, define the experimental setup in section \ref{subsec: datasets} and \ref{subsec: setup} and we validate our methodology in Section \ref{subsec:corr_ana_results}. Then we extend our evaluation with a case study in Section \ref{sec: case-studies}.

\subsection{Datasets}\label{subsec: datasets}
We use three datasets from two domains. From the malware classification domain, we use two complementary datasets, one based on static analysis, and the other on execution behavioral analysis. From image classification, we use a benchmark handwritten digits recognition dataset. We selected these two as representative domains because (a) malware detection is a naturally adversarial domain where adversarial robustness to evasion attacks is expected and (b) image recognition has been heavily explored for evasion attacks in recent adversarial ML literature \cite{Carlini-list}. We describe these datasets next.

\textbf{CuckooTrace (PE Malware).} We collected 40K Windows PE files with 50\% malware (collected from VirusShare~\cite{virusshare}) and the other 50\% benign PEs (collected from a public goodware site~\cite{cnet}). We use 60\% of the dataset as a training set for the target black-box model, 25\% as a training for explanation substitute model, and the remaining 15\% as evasion test. Each sample is represented as a binary feature vector. Each feature indicates the presence/absence of behavioral features captured up on execution of each PE in the Cuckoo Sandbox \cite{cuckoo}. Behavioral analysis of 40K PEs resulted in 1549 features, of which 80 are API calls, 559 are I/O system files, and 910 are loaded DLLs.

\textbf{EMBER (PE Malware).} To assess our framework on complementary (static analysis-based) malware dataset, we use EMBER~\cite{EMBER2018}, a benchmark dataset of malware and benign PEs released with a trained LightGBM with 97.3\% test accuracy. EMBER consists of 2351 features extracted from 1M PEs using a static binary analysis tool LIEF~\cite{lief}. The training set contains 800K samples composed of 600K labeled samples with 50\% split between benign and malicious PEs and 200K unlabeled samples, while the test set consists of 200K samples, again with the same ratio of label split. VirusTotal~\cite{virustotal} was used to label all the samples. The feature groups include: PE metadata, header information, byte histogram, byte-entropy histogram, string information, section information, and imported/exported functions. We use 100K of the test set for substitute model training, and the remaining 100K as our evasion set against the LightGBM pre-trained model and a DNN which we trained. We use version 2 of EMBER. 

\textbf{MNIST (Image).} To further evaluate our framework on image classifiers, we use the MNIST \cite{MNIST} dataset, which comprises
60K training and 10K test images of handwritten digits. The classification task is to identify the digit corresponding to each image. Each 28x28 gray-scale sample
is encoded as a vector of pixel intensities in the interval [0, 1].

\subsection{Models and Setup}\label{subsec: setup}

\textbf{Studied ML Models.}\label{subsec:models}
Across CuckooTrace, EMBER and MNIST, we train $8$ models: Multi-Layer Perceptron (MLP), Logistic Regression (LR), Random Forest (RF), Extra Trees (ET), Decision Trees (DT), Light Gradient Boosting decision tree Model (LGBM), a Deep Neural Network (DNN), and a 2D Convolutional Neural Network (CNN). As in prior work \cite{transferability16,Practical-black-box16}, we choose these models because they are representative of applications of ML across domains including image classification, malware/intrusion detection, and they also complement each other in terms of their architecture and susceptibility to evasion. 

\textbf{Employed Evasion Attacks.} Using the evasion set of each dataset, we craft adversarial samples. For the evasion attack, we consider a threat model where the adversary has no knowledge about the target model, but knows features used to train the model (e.g., API calls for malware classifiers, pixels for image classifiers). More precisely, for CuckooTrace and EMBER we incrementally perturb features of a Malware sample until the model flips its label to Benign. Following previous adversarial sample crafting methods \cite{MalGAN17,ExploreAdvEx18}, we adopt only additive manipulations. For instance, for binary features of CuckooTrace (where 1 indicates presence and 0 indicates absence of an API call), we flip only a 0 to 1. Similar to prior work \cite{SmoothGrad17}, we also respect the allowable range of perturbations for each static feature in EMBER (e.g., file size is always positive). For MNIST, we add a random noise to the background of the image to change the original gray-scale of each pixel without perturbing white pixels that characterize the handwritten digit. The outcome is an adversarial image that is still recognizable by humans, but misclassified by the model. Table II shows the comparison between pre-evasion accuracy and post-evasion accuracy. All models exhibit significant drop in the test accuracy after the feature perturbations. We recall that the main purpose of our analysis is to explore the correlation between a perturbed feature and the misclassification result, regardless of the complexity of the evasion strategy. Thus, our choice of perturbation methods is governed by convenience (e.g., execution time) and effectiveness (i.e., results in evasion). 

\textbf{Employed ML Explanation Methods.} Informed by recent studies \cite{explainEval,explainEval2020} that compare the utility of ML explanation methods, we use LIME \cite{LIME} on CuckooTrace and EMBER, and SHAP \cite{SHAP} on MNIST. More specifically, these studies perform comparative evaluations of black-box ML explanation methods (e.g., LIME \cite{LIME}, SHAP \cite{SHAP}, and LEMNA \cite{LEMNA}) in terms of effectiveness (e.g., accuracy), stability (i.e., similarity among results of different runs), efficiency (e.g., execution time), and robustness against small feature perturbations. On the one hand, these studies show that LIME performs best on security systems (e.g., Drebin+ \cite{Drebin18}, Mimicus+ \cite{LEMNA}). Thus, we employ LIME on the two malware detection systems (i.e., CuckooTrace and EMBER). On the other hand, SHAP authors proposed a ML explainer called ``Deep Explainer'', designed for deep learning models, specifically for image classification. Thus, we use SHAP to explain predictions of a CNN  on MNIST. We note that independent recent studies \cite{explainEval,explainEval2020} suggested that both LIME and SHAP outperform LEMNA \cite{LEMNA}.

\subsection{Correlation Analysis Results} \label{subsec:corr_ana_results}

\begin{table*}[t!]
\centering
\caption{Pre-evasion Accuracy and Post-evasion Accuracy across studied models.}

\begin{tabular}{|c|c|r|r|r|} \hline
{\bf Dataset}&{\bf Model}& {\bf Pre-Evasion }& {\bf Post-Evasion} & {\bf Aggregate}\\ 
{}&{}& {\bf Accuracy}& {\bf Accuracy} & {\bf Evasion Accuracy.}\\

 \hline

CuckooTrace & MLP  & $96\%$  &$6.05\%$ & $89.95\%$\\
CuckooTrace & LR & $95\%$  &$21.75\%$ & $73.25\%$\\
CuckooTrace & RF  & $96\%$ &$18.61\%$ &  $77.39\%$\\
CuckooTrace & DT  & $96\%$ &$7.8\%$ & $88.20\%$\\
CuckooTrace & ET & $96\%$ &$16.27\%$ & $79.73\%$\\ 
EMBER & LGBM  & $97.3\%$ &  $56.06\%$ & $40.94\%$\\
EMBER & DNN & $93$\% & $12.08$\% & $80.92\%$ \\ 
MNIST & CNN & $99.4$\% & $33.1$\%& $66.30\%$ \\  \hline
\end{tabular}

\label{tab:models}

\end{table*}

\begin{table*}[t]
\centering
\caption{High-Correlation Rate and Average Perturbation Error for all models.}

\begin{tabular}{|c|c|r|r|} 
 \hline
{\bf Dataset} & {\bf Model}  & 
{\textbf{High-Correlation Rate}}&
{\textbf{Average Perturbation Error}}\\ \hline

CuckooTrace & MLP                   &   $34.56\%$         &$32.96\%$ \\
CuckooTrace & LR                  &   $34.96\%$      &$38.92\%$\\
CuckooTrace & RF  &  $36.09\%$      &$60.73\%$\\
CuckooTrace & DT   &  $66.85\%$ &$37.86\%$\\
CuckooTrace & ET  &  $33.41\%$   &$54.62\%$\\ 
EMBER & LGBM                  &   $95.03\%$        &$7.47\%$ \\ 
EMBER & DNN                  & $96.72\%$        &$4.89\%$ \\   
MNIST & CNN                  &   $44.31\%$      & $49.40\%$ \\ \hline
\end{tabular}

\label{tab:TER_APE}
\end{table*}


\textbf{Results Overview.} Across all models and the three datasets, the evasion attack scores an average $HCR=55\%$ and $APE=36\%$. Linking back to what these metrics mean, an average $HCR=55\%$ shows that for each model an average of only $55\%$ of the adversarial samples have strong feature-level correlation with their respective perturbations. That entails an average of $45\%$ adversarial samples per-model are loosely correlated with their perturbations. {\em APE} assesses the per-model average number of negative perturbations per sample. Results in Table \ref{tab:TER_APE} suggest a significant rate of negative perturbations are produced by the evasion attack. More precisely, on average across all models around $36\%$ of the perturbations are negative (i.e., they lead the evasion strategy in the wrong direction, by increasing the likelihood of predicting the original label). Next, we expand on these highlights of our findings.

\textbf{\textit{Does evasion imply correlation between perturbations and explanations?}} Although an evasion attack can achieve a seemingly high aggregate evasion rate (e.g., as high as $94\%$ accuracy drop on MLP on CuckooTrace), we notice that, the correlation between each single feature perturbation and a misclassification is not guaranteed. In fact, averaged across models, $45\%$ of the crafted adversarial samples have low-correlated perturbations more than high-correlated ones ($PSPP(x')<0.5$), suggesting that almost 1 in 2 adversarial samples suffers from weak correlation between post-evasion explanations and pre-evasion perturbations. As a result, counting in such samples in the aggregate evasion rate would essentially give false sense of the effectiveness of an attack strategy at the granularity of each feature perturbation. We underscore that such insights would not have been possible to infer without the high-fidelity correlation analysis. In summary,  these results confirm that {\em not all evasive predictions of an adversarial sample are correlated with the performed feature manipulations}.

\begin{figure}[htb!]
    
    \centering
    \includegraphics[width=\columnwidth]{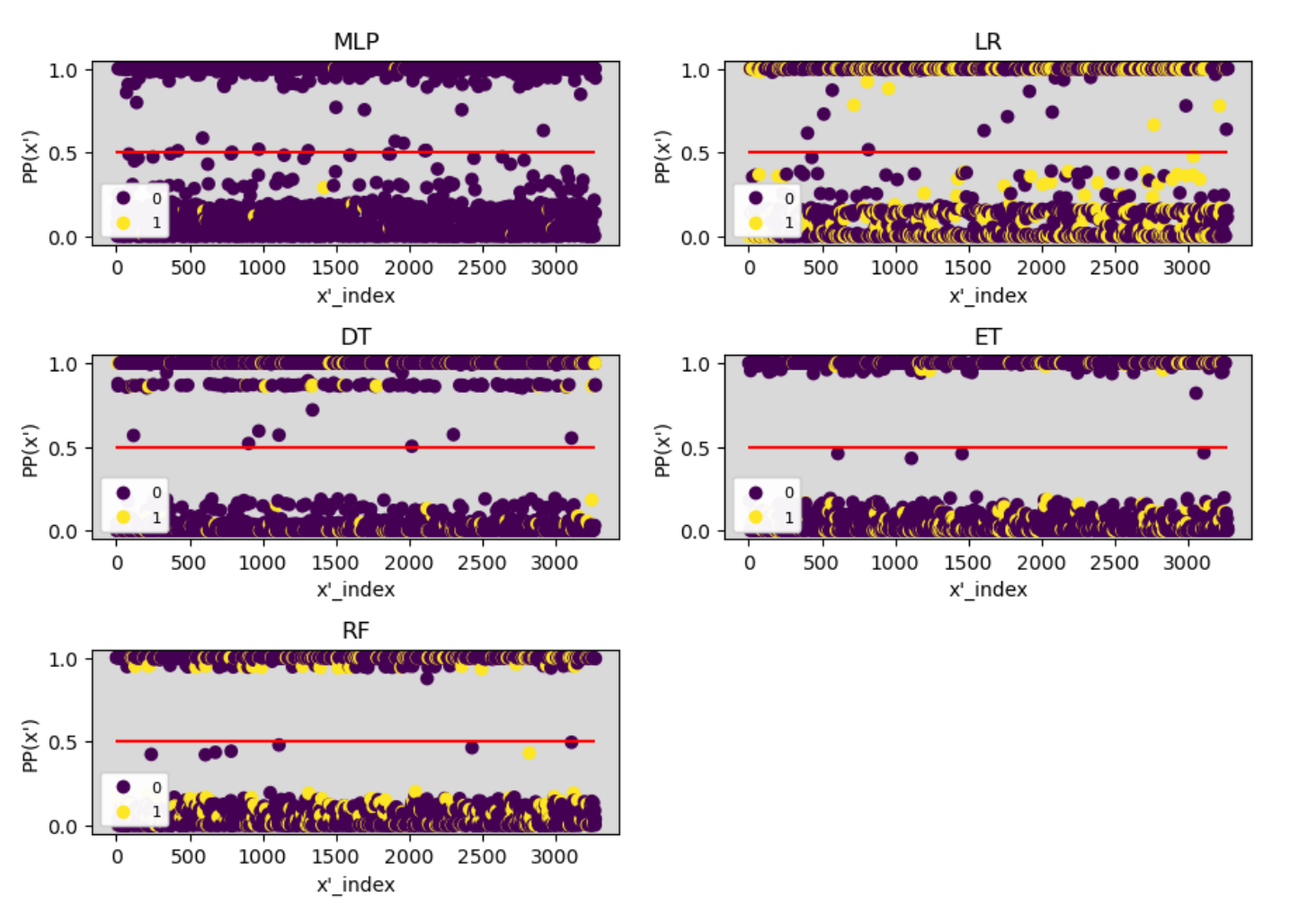}
    \caption{Distribution of Perturbation Precision (PSPP) values of each adversarial sample across models on CuckooTrace.}
    \label{fig:PP1}

\end{figure}

\textbf{\textit{Visualizing the Per-Sample Perturbation Precision.}} Figure \ref{fig:PP1} shows the distribution of $PSPP$ values of all crafted malware samples of CuckooTrace  across the 5 models. In the figure, we use the shorthand $PP$ instead of $PSPP$ in the $y$-axis and we refer to the index of each sample in $x$-axis. The true prediction of each malware sample is $f_b(x')=1$, while the evasive prediction is $f_b(x') = 0$ (i.e., adversarial malware sample is misclassified as benign). The red line in the middle represents the threshold $\tau$ that decides whether the adversarial sample has more positive perturbations or more negative ones. In almost all the plots, we notice the occurrence of a significant number of low-correlation adversarial malware samples ($PP(x')<\tau$) that evaded the classifier (purple circles below the red line). Once again, these findings suggest that the evasion attack results in a high number of negative perturbations. However, despite the low number of positive perturbations for these samples, the evasion is still successful. This result goes along with our previous finding that high evasion aggregate accuracy can have low correlation with  performed perturbations. Thus, even for a successful evasion the perturbations at the feature-level can apparently be ineffective. 
  
It is noteworthy that Figure \ref{fig:PP1} exhibits an unusual behavior. In particular, some crafted samples with a true prediction $f_b(x')=1$ (yellow circles) appear to have high Perturbation Precision, $PP(x')>\tau$, despite the failure to flip the model's prediction from $1$ to $0$.  This is especially true for models: LR, DT, ET and RF. On one hand, this observation suggests  that even a small number of negative perturbations ($PP(x')<\tau$) may affect the final outcome of the evasion. On the other hand, it suggests potential limitations of Black-Box ML explanation methods in terms of accuracy and stability between different runs. More discussion is provided about this in Section \ref{sec:discussion}.

\textbf{\textit{What do correlation analysis results suggest across different classification tasks and model architectures?}} While our results so far strongly suggest the importance of post-evasion correlation analysis for an in-depth assessment of an evasion attack strategy, we also observe that $HCR$ and $APE$ values vary across studied domains (malware, image), model architectures, and feature representations (static, dynamic). This variation speaks to the sensitivity of different domains, models, and feature values to adversarial feature perturbations with implications on robustness and dependability in the face of individual feature perturbation. In fact, ML models trained on EMBER (LGBM and DNN) showed acceptably low rate of negative perturbations (i.e., $APE = 6\%$ on average) and a high rate of samples with highly-correlated perturbations (i.e., $HCR = 96\%$ on average). This suggests that static features of Windows PE malware are more sensitive to a feature perturbation considering the higher rate of negative perturbations on dynamic features in CuckooTrace. In terms of comparison between different domains and different ML models, despite the high evasion rate at sample-level, almost all ML models showed some robustness at the level of a single feature perturbation. Most importantly, RF on CuckooTrace showed the highest robustness since more than $60\%$ of the overall feature perturbations are negative which suggest that they did  not contribute in the misclassification decision. ET on CuckooTrace ($APE = 54\%$) and CNN on MNIST ($APE = 49\%$) showed lower robustness than RF, but higher than the other models.

 \textbf{Summary.} Our results suggest that aggregate evasion accuracy is inadequate to assess the efficacy of perturbation attack strategy. Our findings also validate that explanation-guided correlation analysis plays a crucial role in diagnosing aggregate evasion rates to winnow high-correlation adversarial samples from low-correlation ones for precise feature-level assessment of evasion accuracy.

\section{Case Study}\label{sec: case-studies}
We now present a case study that demonstrates the application of our correlation analysis framework on explanation-guided evasion strategy.

\subsection{Explanation-Guided Evasion Strategy}\label{subsec: EGE}
The correlation analysis results showed that,  while an evasion strategy may result in an evasive adversarial sample, at the granularity of a single feature perturbation it may produce a considerable number of \textit{negative perturbations}. In other words, from the adversary's standpoint, the correlation analysis can be leveraged towards more accurate evasion strategy that significantly minimizes negative perturbations. In the following, we explore the potential of explanation methods to guide a more effective evasion strategy.

\begin{figure}[t!]
    
    \centering
    \includegraphics[width=\columnwidth]{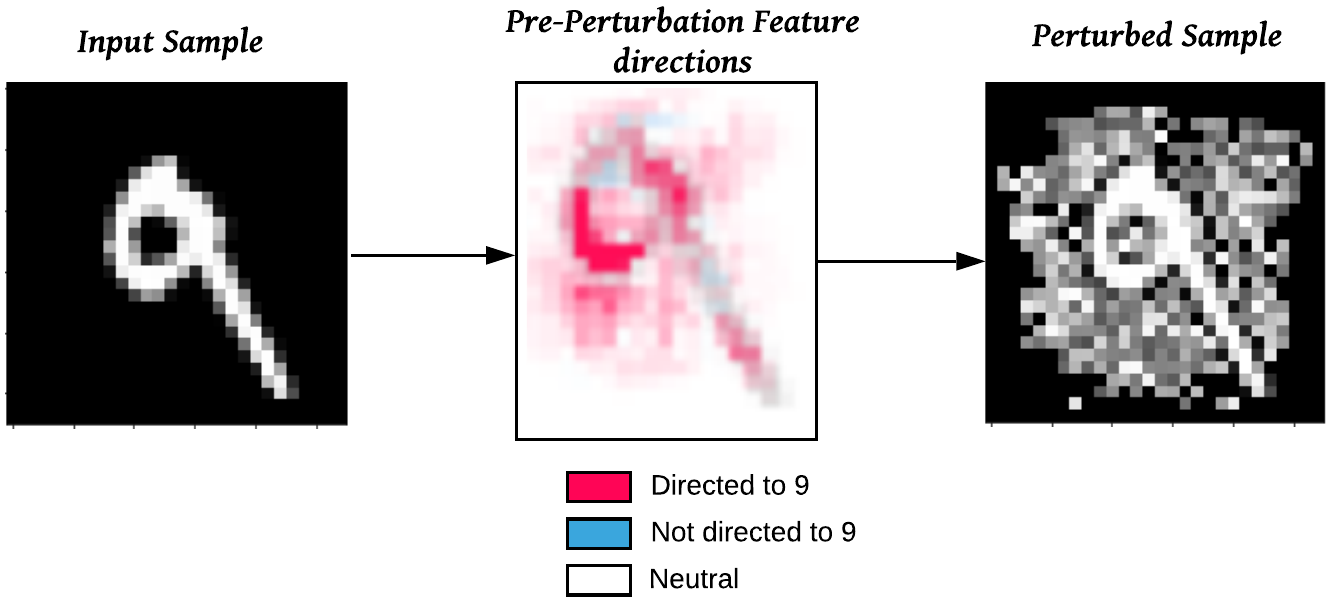}
    \caption{An example of explanation-guided feature perturbations on an input sample from MNIST.}
    \label{fig:ege}

\end{figure}

In this case study, we demonstrate how an adversary leverages ML explanation methods to examine pre-perturbation predictions before making feature manipulations. In particular, the \textit{pre-perturbation feature directions} reveal \textit{positive features} that significantly contribute to the true prediction (pink pixels in Figure \ref{fig:ege}). Intuitively, positive features are strong candidates for perturbations, while \textit{negative features} (blue pixels in Figure \ref{fig:ege}) need not be perturbed since they are already directed away from the true label, which is in favor of the adversary's goal. \textit{Neutral features} (white pixels in Figure \ref{fig:ege}) are also not candidates for perturbations since they have no effect on the original label decision. We note that in this case study we consider all positive features (i.e., pink pixels) as candidates for perturbation regardless of the color intensity that represents its explanation weight. In Figure \ref{fig:ege}, some positive pixels ($w_i > 0$) with a low explanation weight ($w_i \sim 0$) are almost neutral (i.e., closer to the white color) but still perturbed since they are directed to the true label. Using the same experimental setup, we enhance the evasion strategies used on the three datasets with \textit{explanation-guided pre-perturbation feature selection}. Then, we measure changes to \textit{post-evasion accuracy}, \textit{HCR}, and \textit{APE} for all studied models.

\begin{table*}[htb!]
\centering
\caption{Post-Evasion Accuracy, HCR, and APE values across studied models using explanation-guided evasion.}

\begin{tabular}{|c|c|r|r|r|} \hline
{\bf Dataset}  &{\bf Model}  & 
{\textbf{Post-Evasion Accuracy}}&
{\textbf{HCR}}&
{\textbf{APE}}\\ \hline

CuckooTrace & MLP &$0\%$ & $92.03\%$ &$14.22\%$\\
CuckooTrace  & LR &$0\%$ & $96.25\%$  &  $6\%$\\
CuckooTrace & RF  &$0\%$& $48.53\%$ &$51.31\%$\\
CuckooTrace & DT   &$1.41\%$& $98.84\%$ & $3.41\%$\\ 
CuckooTrace & ET  &$0\%$ & $48.11\%$  &$1.21\%$\\ 
EMBER & LGBM & $27.16\%$& $99.58\%$ &$2.69\%$ \\ 
EMBER & DNN & $11.7\%$& $97.8\%$ &$2.71\%$\\  
MNIST & CNN &$24.67\%$& $64.5\%$ &$43.4\%$\\  \hline 
\end{tabular}
\label{tab:TER_APE2}

\end{table*}
\subsection{Impact of Explanation-Guided Evasion}
Results in Table \ref{tab:TER_APE2} suggest overall improvement not only in aggregate evasion accuracy, but also in the correlation strength between evasion explanations and individual feature perturbations. Comparing the ``Post-Evasion Accuracy'' columns of Tables \ref{tab:models} and \ref{tab:TER_APE2}, using  explanation-guided evasion strategy post-evasion accuracy drops for all studied models, with an average per-model drop of $13.4\%$ (which translates to the same percentage of improvement in aggregate evasion accuracy). Interestingly, in 4 out of the 5 models in CuckooTrace, post-evasion accuracy drops to zero, with up to $21\%$ drop in post-evasion accuracy for models such as LR. We note that the eventual complete evasion in almost all models in CuckooTrace is most likely attributed to the binary nature of the features, where the explanation-guided feature selection filters out negative features and leaves only positive features that are flipped with just one perturbation. Comparing the $HCR$ columns of Tables \ref{tab:TER_APE2} and \ref{tab:TER_APE}, we notice an increase in $HCR$ for all studied models. On average, $HCR$ increased by $27\%$ per-model, which shows the positive utility of the pre-perturbation explanations that guided the evasion strategy to perturb positive features instead of negative ones. Again, comparing the $APE$ columns of Tables \ref{tab:TER_APE2} and \ref{tab:TER_APE}, we notice a significant drop in $APE$, with an average per-model decrease of $20\%$, which indicates a decrease in the number of \textit{negative perturbations}. Better performance in terms of post-evasion accuracy is also observed for all studied target models. 

We note that although we perturb only \textit{positive features}, in the $APE$ column of Table \ref{tab:TER_APE2} all values are still non-zero. Ideally, the explanation method would guide the perturbation strategy to perform only positive perturbations and make no mistaken perturbations. Nevertheless, we still observe a minimal percentage of \textit{negative perturbations} due to the inherent limitations of the accuracy and stability of explanations by LIME and SHAP, which is also substantiated by recent studies \cite{explainEval,explainEval2020} that evaluated LIME and SHAP among other ML explanation methods. We will expand on limitations of ML explainers in Section \ref{sec:discussion}.
 
 \textbf{Summary.} This case study suggests that when used on top of existing feature perturbation methods, an explanation-guided feature selection strategy leads to more effective evasion results both in terms of aggregate evasion accuracy and effectiveness at the level of each feature manipulation.

\section{Discussion and Limitations}\label{sec:discussion}
Recent studies \cite{explainEval,explainEval2020} have systematically compared the performance of ML explanation methods especially on security systems. In addition to general evaluation criteria (e.g., explanation accuracy and sparsity), Warnecke et al. \cite{explainEval} focused on other security-relevant evaluation metrics (e.g., stability, efficiency, and robustness). Fan et al. \cite{explainEval2020} also proposed a similar framework that led to the same evaluation results. Next, we highlight limitations of LIME \cite{LIME} and SHAP \cite{SHAP} based on \textit{accuracy} (degree to which relevant features are captured in an explanation), \textit{stability} (how much explanations vary between runs), and \textit{robustness} (the extent to which explanations and prediction are coupled).

\textbf{Limitations of Explanation Methods.} While LIME and SHAP produce more accurate results compared with other black-box explanation methods (e.g., DeepLIFT \cite{DeepLIFT}, LEMNA \cite{LEMNA}), the \textit{accuracy} of the explanation may vary across different ML model architectures (e.g., MLP, RF, DT, etc), and across different ML tasks/datasets (e.g., CuckooTrace, EMBER, and MNIST). For instance, the inherent linearity of LIME's approximator could negatively influence its accuracy and stability in explaining predictions of complex models such as RF and ET.  More importantly, like all learning-based methods, LIME and SHAP are sensitive to non-determinism (e.g., random initialization, stochastic optimization) which affect their \textit{stability} between different runs. In other words, it is likely to observe a slight variation in the output of multiple runs performed by the same explanation method using the same input data. In fact, we observed that the average difference between Shapley values (i.e., feature importance weights) returned by SHAP is around $1\%$ over 100 runs on the same MNIST sample. Such variation in ML explanation outputs might partly explain some of the unexpected results of our explanation-guided analysis that we noted in Section \ref{subsec:corr_ana_results}.

\textbf{Vulnerability of Explanation Methods.} Another issue worth considering is \textit{robustness} of ML explanation methods against adversarial attacks. Studies \cite{exp_Rob1,exp_Rob2,EXML-UnderFire20} have demonstrated that the explanation results are sensitive to small systematic feature perturbations that preserve the predicted label. Such attacks can potentially alter the explanation results, which might in effect influence our explanation-guided analysis. Consequently, our analysis may produce potentially misleading results for correlation metrics such as HCR and APE. 
 In light of the utility of ML explanations we demonstrated in this work, we hope that our framework can be instantiated for adversarial perturbations performed in the problem-space. We note, however, that there needs to be careful consideration in mapping the units of adversarial perturbations in problem space manipulations (e.g., the organ transplant notions proposed in~\cite{ProblemSpace-20}) to the metrics we proposed in Section \ref{sec: approach}. Finally, we note that vulnerability to adversarial attacks is a broader problem for any ML model, and progress in defense against attacks such as adversarial examples will potentially inspire and inform stronger robustness properties for ML explanation methods.
\section{Related Work}\label{sec: related}
 We discuss related work focusing on evasion attacks and ML explanation methods.

Comparing how evasion is assessed, our approach is complementary to prior work ~\cite{FGSM,BIM,PGDM,CW,EndtoEnd18,GrossePMBM17,ExploreAdvEx18}, which rely on aggregate evasion accuracy.
Next, we shed light on relevant prior work focusing on evasion of image classification and malware/intrusion detection systems.

\textbf{Image Classification.} Several evasion methods have been proposed for image classification tasks~\cite{Carlini-list}. Some of the most notable ones are: Fast Gradient Sign Method (FGSM)~\cite{FGSM}, Basic Iterative Method (BIM)~\cite{BIM}, Projected Gradient Descent (PGD) method~\cite{PGDM}, and Carlini \& Wagner (CW) method~\cite{CW}. 

\textbf{Windows Malware.} Al-Dujaili et. al.~\cite{SLEIPINIR18} leverage saddle-point formulation towards an adversarial training that incorporates adversarial samples to train models robust against gradient-based attacks such as FGSM\cite{FGSM} and BIM\cite{BIM}. Kolosnjaji et. al.~\cite{MalConvEvade18} proposed a gradient-based attack against MalConv~\cite{malconv18} by appending bytes (in the range 2KB-10KB) to the overlay of a PE. As a follow-up to~\cite{MalConvEvade18}, Demetrio et. al.~\cite{ExpVulns19} extended the adversarial sample crafting method by demonstrating the feasibility of evasion by manipulating 58 bytes in the DOS header of PE. Suciu et. al.~\cite{ExploreAdvEx18} explored FGSM~\cite{FGSM} to craft adversarial samples against MalConv \cite{MalConvEvade18} by padding adversarial payloads between sections in a PE provided that there is space to perform padding. Hu and Tan \cite{MalGAN17} train a substitute model using the GAN framework to fit a black-box malware detector trained on API call traces of Windows PEs. In a follow-up work \cite{RNN-Black-Box18}, they use the GAN-based substitute model training for a recurrent neural network.

Other works utilize reinforcement learning to evade malware classifiers. For instance, Anderson et. al.~\cite{Anderson-Reinf17} aim to limit perturbations to a select set of transformations that are guaranteed to preserve semantic integrity of a sample using reinforcement learning. Additionally, Apruzzese et. al. ~\cite{Apruzzese2020DeepRA} propose a method that generates realistic attack samples that can evade botnet detection through deep reinforcement learning. The generated samples can be used for adversarial learning. Rosenberg et. al.~\cite{APIBlackBoxEvade18} adopt the Jacobian-based feature augmentation method introduced in \cite{Practical-black-box16} to synthesize training examples to inject fake API calls to PEs at runt-time. Using the augmented dataset, they locally train a substitute model to evade a target RNN black-box malware detector based on API call features.

\textbf{Android Malware.} Similar to \cite{MalGAN17} and \cite{RNN-Black-Box18}, Grosse et. al. ~\cite{GrossePMBM17} demonstrate evasion by adding API calls to malicious Android APKs. Yang et. al.~\cite{Android-YangK0G17} explore semantic analysis of malicious APKs with the goal of increasing resilience of Android malware detectors against evasion attacks. Recently, Pierazzi et. al.~\cite{ProblemSpace-20} take a promising step towards formalization of the mapping between feature space and problem space on the Drebin classifier~\cite{Drebin14} for Android malware.

\textbf{PDF Malware.} Srndic et. al.~\cite{PDF-Evasion-SrndicL14} are among the first to demonstrate vulnerabilities of deployed PDF malware detectors using constrained manipulation with semantic preservation. Xu et. al.~\cite{PDF-XuQE16} use genetic algorithms to manipulate ASTs of malicious PDFs to generate adversarial variants while preserving document structure (syntax).

\textbf{Explanation Methods.} 
Several ML explanation methods have been proposed by prior work \cite{LIME,LEMNA,SHAP,DeepLIFT}. 
In a black-box setting, LIME \cite{LIME}, Anchors \cite{Anchor18}, SHAP \cite{SHAP}, and \cite{Rule-based-Exp} are among the typical class of black-box explanation methods that rely on local approximation. Other methods (e.g., \cite{Meaningful-perturb17,Erasure16}) use input perturbation approaches by monitoring prediction deltas. DeepLIFT \cite{DeepLIFT} explains feature importance with respect to a reference output, while white-box explanation techniques (e.g., \cite{Whitebox-exp13,Whitebox-exp14,SmoothGrad17}) use gradient-based feature importance estimation. Recent studies \cite{explainEval,explainEval2020,exp_Rob1,exp_Rob2} explore the utility of explanation methods across criteria such as accuracy, stability, and robustness of explanations. A more recent interesting application is the use of SHAP signatures to detect adversarial examples in DNNs~\cite{EG-Evasion-Detection20}.  

 In summary, Prior works typically rely on comparing the pre-perturbation accuracy and post-perturbation accuracy in order to evaluate ML evasion attacks, which lacks deeper diagnosis of the attack's success. In this paper, we add complementary suite of metrics in order to map a single feature perturbation with its contribution to evasion.
\section{Conclusion}\label{sec: concl}
We introduced the first explanation-guided methodology for the diagnosis of ML evasion attacks. To do so, we use feature importance-based ML explanation methods to enable high-fidelity correlation analysis between pre-evasion perturbations and post-evasion prediction explanations. To systematize the analysis, we proposed and evaluated a novel suite of metrics. Using image classification and malware detection as representative ML tasks, we demonstrated the utility of the methodology across diverse ML model architectures and feature representations. Through a case study we additionally confirm that our methodology enables evasion attack improvement via pre-evasion feature direction analysis.

\section*{Acknowledgments}
We thank our shepherd Giovanni Apruzzese and the anonymous reviewers for their insightful feedback that immensely improved this paper.
\bibliographystyle{plain}
\bibliography{main}

\end{document}